\documentclass[journal,a4paper]{IEEEtran} 
\IEEEoverridecommandlockouts
\usepackage{cite}
\usepackage{amsmath,amssymb,amsfonts}
\usepackage{algorithmic}
\usepackage{graphicx}
\usepackage{textcomp}
\usepackage{siunitx}
\usepackage{xcolor}
\usepackage{hyperref}

\providecommand{\subcaption}[1]{#1}

\makeatletter
\let\old@ps@headings\ps@headings
\let\old@ps@IEEEtitlepagestyle\ps@IEEEtitlepagestyle
\def\confheader#1{%
\def\ps@IEEEtitlepagestyle{
\old@ps@IEEEtitlepagestyle
\def\@oddhead{\strut\hfill#1\hfill\strut}
\def\@evenhead{\strut\hfill#1\hfill\strut}
}
\ps@headings
}
\makeatother
\confheader{
\small{Accepted and presented at the 13th RSI International Conference on Robotics and Mechatronics (ICRoM 2025)} 
}
\usepackage[pscoord]{eso-pic}

\newcommand{\centerfooter}[2][0.025]{%
  \AddToShipoutPictureFG*{%
    \put(\LenToUnit{0.5\paperwidth},\LenToUnit{#1\paperheight}){%
      \makebox[0pt][c]{#2}%
    }%
  }%
}

\centerfooter{\textbf{\small{Accepted and presented at the 13th RSI International Conference on Robotics and Mechatronics (ICRoM 2025)}}}

\def\BibTeX{{\rm B\kern-.05em{\sc i\kern-.025em b}\kern-.08em
    T\kern-.1667em\lower.7ex\hbox{E}\kern-.125emX}}
\begin{document}

\title{Deep Reinforcement Learning for Adaptive Gain Tuning in Control of Teleoperation Manipulators with Joint Flexibility and Time-Varying Delays\\
}

\author{
    \IEEEauthorblockN{
        1\textsuperscript{st} Armin Attarzadeh,
        2\textsuperscript{nd} Mohammad Ali Ghaemifar,
        3\textsuperscript{rd} Alireza Khanzadeh,
        4\textsuperscript{th} Soheil Ganjefar
    }

    \IEEEauthorblockA{
        \textit{Department of Electrical Engineering, Iran University of Science and Technology (IUST)}\\
        Tehran, Iran\\
        Email: armin\_attarzadeh, m\_ghaemifar,  alireza\_khanzadeh75@elec.iust.ac.ir, s\_ganjefar@iust.ac.ir
    }
}

\maketitle

\begin{abstract}
Bilateral teleoperation systems that include joint flexibility better reflect real robotic systems used in surgery, space, and rehabilitation. However, joint flexibility together with time-varying communication delays makes it difficult to maintain stable and coordinated motion between the master and slave robots. To address this, we propose a hybrid control method that combines a stable Proportional-plus-Damping (P+d) controller with a model-free deep reinforcement learning agent based on the Twin Delayed Deep Deterministic Policy Gradient (TD3) algorithm. The P+d controller provides basic stability under bounded delays, while the learning agent adjusts and tunes the remote-side proportional and damping gains in real time to reduce vibrations and improve tracking. Stability is guaranteed for bounded time-varying delays using Lyapunov–Krasovskii analysis. The approach provides a practical solution for teleoperation systems facing both joint flexibility and uncertain network delays.
Code is publicly available at
\href{https://github.com/ArminAttarzadeh/DRL-Controller-Gain-Tuner}{github.com/ArminAttarzadeh/DRL-Controller-Gain-Tuner}
\end{abstract}


\section{Introduction}
A teleoperation manipulator system consists of a human operator, a local master robot, a communication channel, and a remote slave robot. The remote robot operates within a distant environment. Teleoperation enables human manipulation and sensory capabilities in remote or hazardous locations. The system is designed to ensure stable and transparent interaction between the master and slave robots.
Teleoperation and robotic control technologies have facilitated substantial advancements in domains that require remote manipulation, including space exploration \cite{b1}, rehabilitation \cite{b2}, and surgery \cite{b3}. These application areas underscore the need for advanced teleoperation systems capable of managing complex and high-risk procedures. Commercial platforms such as the da Vinci surgical system \cite{b4} and the Sina teleoperation system \cite{b5} exemplify practical implementations, highlighting the continuous development of teleoperation technologies.
Robotics research originated with Goertz's development of the first master–slave manipulator in the 1940s, which established bilateral teleoperation. In subsequent decades, researchers addressed key challenges in remote robotic interaction, including stability, delay compensation, and transparency. These efforts encompassed supervisory and software-based control, passivity theory, scattering methods, and Internet-based architectures \cite{b6}. These issues have motivated the control theoretic research in teleoperation over the past decades.
The scattering transformation, introduced by Anderson and Spong \cite{b7} in the late 1980s, has played a key role in teleoperation system control. However, most methods based on this approach experience position drift, with only a few exceptions \cite{b8}. To overcome this, Chopra et al. \cite{b9} developed adaptive control schemes that reduce position drift without using scattering variables. Similarly, Nuño et al. presented another adaptive strategy that synchronizes the positions of master and slave manipulators even when constant time delays are present, and later provided a corrected stability proof \cite{b10}.
Recent research has focused on overcoming stability and transparency challenges in bilateral teleoperation systems operating under constant communication delays. For example, Ganjefar et al. \cite{b32} proposed the
utilization of wave-variables and the Smith predictor as
strategies to minimize the impact of delays. Venkateswaran and Qu \cite{b11} introduced a passivity-shortage-based control framework. This framework extends conventional passivity theory to systems with constant delays. Unlike the classical scattering or wave-variable methods, their design uses a negative-feedback interconnection of passivity-short systems. This approach guarantees L² stability even when the delay violates strict passivity \cite{b11}. Lu et al. \cite{b12} proposed an adaptive bilateral controller based on the position-error-based (PEB) structure to handle uncertain dynamics and constant communication delay. Through Lyapunov analysis, they proved global stability and verified the controller on a 2-DOF master–slave manipulator in simulation. Furthermore, the adaptive law compensated for parameter uncertainty and achieved accurate position tracking \cite{b12}. Wang et al. \cite{b13} combined terminal sliding-mode control with neural-network adaptation to obtain finite-time convergence in a force-feedback teleoperation system under constant delay. The Lyapunov-based proof guaranteed boundedness of all signals and ensured that master–slave tracking errors vanished in finite time \cite{b13}. Yang et al. \cite{b14} developed a wave-variable compensator for a three-channel (3CH) bilateral teleoperation system with constant communication delay. They reformulated the 3CH architecture into a two-port passive network. Then, they introduced two wave-variable compensators with energy reservoirs. This approach minimized the bias that normally degrades transparency \cite{b14}. 
Uyulan \cite{b15} introduced a robust passivity-based nonlinear controller that incorporates high-order sliding-mode observers to address load disturbances and communication delays. This method preserved passivity and transparency by compensating for disturbances using super-twisting observers. In practical teleoperation systems, time-varying delays are inevitable due to fluctuating network bandwidth, routing changes, and signal transmission uncertainty. Uyulan et al. tackled this challenge by proposing a robust passivity-based nonlinear controller equipped with a super-twisting sliding-mode observer and a disturbance observer to mitigate the effects of variable delays and load disturbances, ensuring stable coordination between master and slave manipulators \cite{b15}. Building on this robustness perspective, Chang et al. \cite{b16} introduced a robust adaptive sliding-mode control scheme that employs Lyapunov–Krasovskii functionals and LMI-based delay bounds to guarantee stability under bounded time-varying delays, validated through experimental results. Chen et al. \cite{b17} introduced a delay-adaptive synchronization controller that uses position-only communication, a velocity-feedback filter, and a radial basis function neural network. This approach addresses asymmetric time-varying delays and compensates for system uncertainties. Similarly, Lu et al.\cite{b12} proposed an adaptive control method for teleoperation systems with uncertain dynamics, focusing on parameter adaptation to maintain stability in the presence of delays and modeling errors. Chen et al. \cite{b18} advanced from adaptation to prediction by introducing a Smith predictive controller for space teleoperation. Their approach uses a delay estimation model to forecast communication latency and maintain transparency despite significant delay variations. Building on this, Ebrahimian et al. \cite{b19} integrated model-mediated environment force prediction with an artificial neural network that predicts the operator’s future motion, enabling real-time synchronization between master and slave systems even when delays are unpredictable. More recently, Dao et al. \cite{b20} addressed the same problem using a rise-based integral reinforcement learning (IRL) framework, where an actor–critic agent learns optimal control laws for systems with variable delays and disturbances, achieving both coordination and robustness. Meanwhile, Patiño et al. \cite{b21} employed neural network compensators in a skid-steer teleoperation system, where damping injection and learned interaction models jointly ensured stable motion tracking despite nonlinear time-delay effects. In a related approach, Xia et al. \cite{b22} reframed the problem from a learning standpoint by introducing a belief-state actor–critic (BSAC) algorithm that explicitly models action and observation delays within an agent-based space teleoperation framework, effectively mitigating both constant and random time lags. In a different teleoperation configuration involving one master and three slave manipulators, Khanzadeh et al. \cite{b31} introduced an adaptive synchronization controller designed to maintain coordination and stability under time-varying delays and formation changes.
Incorporating joint or link flexibility into the teleoperation framework aligns system dynamics more closely with those of real-world manipulators. This increased realism introduces additional nonlinearities, vibrations, and underactuation, which complicate the control problem, particularly in the presence of time delays. To address these challenges, several recent control strategies have been developed. Nuño et al. \cite{b23} developed adaptive and P+d controllers for teleoperation systems with joint flexibility and time delays, proving that both can ensure stable motion and asymptotic position tracking. By explicitly modeling joint elasticity, they showed that flexible manipulators more accurately represent real teleoperation dynamics while maintaining global stability through Lyapunov–Krasovskii analysis. Xu et al. \cite{b24} developed a TD3-based adaptive controller for a circulating cooling water system, showing that TD3 algorithms outperform PID, fuzzy PID, and DDPG methods in nonlinear environments. Their use of a reference trajectory model improved convergence and reduced oscillations, demonstrating the potential of deep reinforcement learning for systems with slow and uncertain dynamics. In parallel, Yin et al. \cite{b25} utilized a TD3 reinforcement learning controller for the vector control of permanent magnet synchronous motors (PMSM). By replacing the conventional PI controller within the current loop with a learned policy, they achieved faster response and reduced tracking error under variable operating conditions. Their results confirmed that DRL can effectively tune control gains online in complex continuous-action systems, a principle extendable to flexible manipulators. Barjini et al. \cite{b26} created a hybrid approach for flexible robotic arms that blends a physics-based controller with a learning-based motion planner. The controller keeps the system stable, while the planner uses soft actor–critic (SAC) methods to find movements that reduce unwanted vibrations. This combination led to better control and less shaking, as shown in lab tests on a flexible robot arm. Despite these advances, most existing studies focus on either compensating for time-varying delays or addressing joint flexibility. Few studies tackle both challenges simultaneously within a unified framework. Deep reinforcement learning shows promise in managing nonlinear and uncertain dynamics. However, its integration with bilateral teleoperation systems that include elastic joints and variable communication delays remains largely unexplored. Traditional adaptive and neural controllers still rely on manual gain tuning or precise model knowledge. This reliance limits their adaptability in unpredictable networked environments.

\begin{figure}[!t]
    \centering
    \includegraphics[width=0.8\columnwidth]{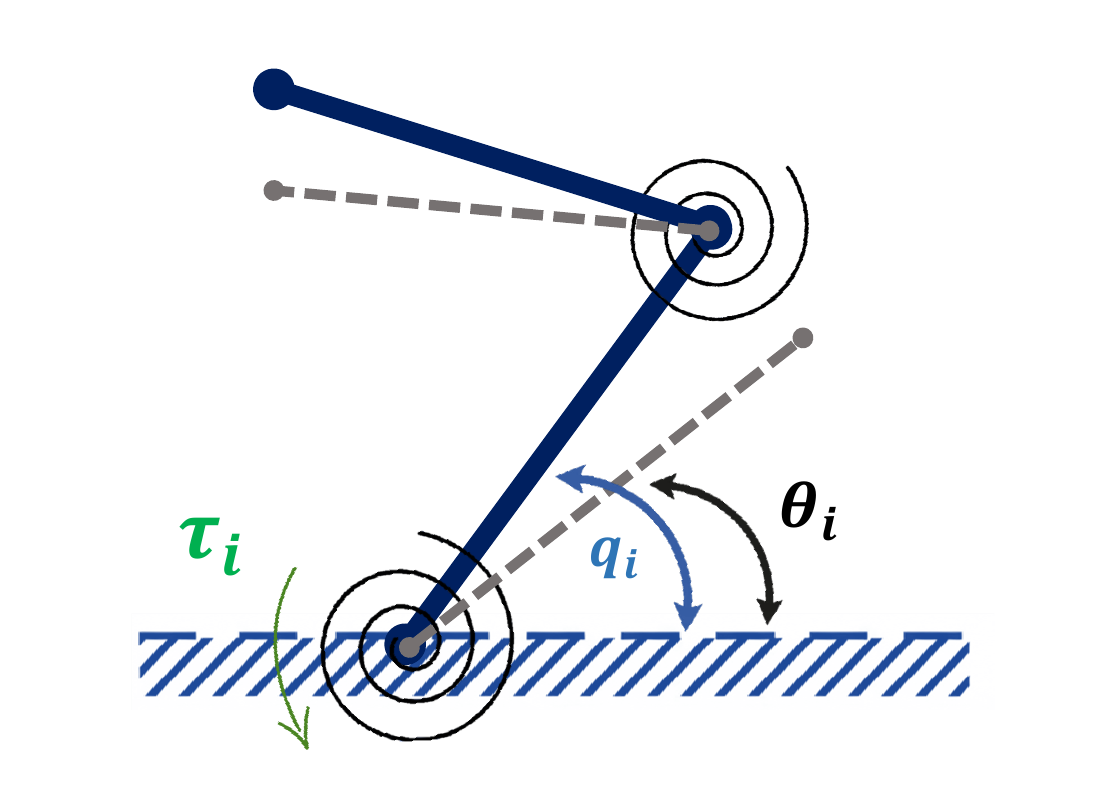}
    \caption{Schematic of the 2-DOF flexible joint manipulator.}
    \label{fig:robot}
\end{figure}

\section{Dynamic Model of the Bilateral Teleoperation System}

Firstly, gravity is eliminated in both the local and remote robots. After, we consider the flexible teleoperation system according to Fig.1, its non-linear dynamic behavior of the remote manipulator is given by \cite{b27,b28}:
\begin{align}
M_r (q_r ) \ddot{q}_r + C_r (q_r, \dot{q}_r ) \dot{q}_r + S_r [q_r - \theta_r] &= -\tau_e \label{eq:1a} \\
J_r \ddot{\theta}_r + S_r [\theta_r - q_r] &= \tau_r \label{eq:1b}
\end{align}
The local is modeled as a manipulator with rigid joints. Its nonlinear behavior is governed by \cite{b10}:
\begin{equation}
M_l (q_l ) \ddot{q}_l + C_l (q_l, \dot{q}_l ) \dot{q}_l = \tau_h - \tau_l \label{eq:2}
\end{equation}
Where \( i \in \{l, r\} \) for local and remote manipulators, \( q_i, \dot{q}_i, \ddot{q}_i \in \mathbb{R}^n \) denote the joint positions, velocities, and accelerations, and \( \theta_r \in \mathbb{R}^n \) is the remote joint (motor) position, \( M_i \in \mathbb{R}^{n \times n} \) are the inertia matrices. \( C_i \in \mathbb{R}^{n \times n} \) represent the Coriolis and centrifugal effects, \( J_r \in \mathbb{R}^{n \times n} \) is a constant diagonal matrix representing the remote actuator moments of inertia, \( S_r \in \mathbb{R}^{n \times n} \) is a constant diagonal and positive definite matrix that contains the remote joint stiffness, \( \tau_i \in \mathbb{R}^n \) are the control signals, and \( \tau_h \in \mathbb{R}^n \), \( \tau_e \in \mathbb{R}^n \) are the operator and environment torques. This model has the following properties \cite{b29}.
\begin{itemize}
\item Property 1. The inertia matrices \( M_i (q_i ) \) are positive definite and bounded.
    \item Property 2. The matrix \( \dot{M}_i (q_i ) - 2C_i (q_i, \dot{q}_i ) \) is skew-symmetric, \( \dot{M}_i (q_i ) = C_i (q_i, \dot{q}_i ) + C_i^T (q_i, \dot{q}_i ) \).
\end{itemize}

\section{Proposed Control Strategy}

We introduce a hybrid control approach that combines a stable proportional plus damping injection ($P+d$) controller with a deep reinforcement learning (DRL) agent as an adaptive gain tuner. The $P+d$ controller ensures baseline stability, while the DRL agent continuously adjusts the proportional ($K$) and damping ($B$) gains in real time. This approach suppresses flexible-joint vibrations and maintains precise tracking despite unpredictable, time-varying communication delays, delivering greater robustness and adaptability compared to fixed-gain static controllers.

\subsection{Proportional Plus Damping Injection (\texorpdfstring{$P+d$}{P+d}) Controller}\label{AA}

The $P+d$ controller serves as a core stabilizing element for bilateral teleoperation systems. This architecture has been shown in prior studies to ensure global stability for nonlinear teleoperators, establishing position tracking even in the presence of joint flexibility and variable time-varying delays. The specific control scheme implemented assumes a rigid local manipulator and a remote manipulator with flexible joints. Under this scheme, velocities and position errors remain bounded, and in the absence of external forces, exact position tracking is guaranteed. The formal stability guarantee for this control structure will be discussed in Section IV.

For the local (rigid) manipulator, feedback utilizes the link position \(q_l(t)\) and link velocity \(\dot{q}_l(t)\). For the remote (flexible) manipulator, only the joint (motor) position \(\theta_r(t)\) and velocity \(\dot{\theta}_r(t)\) are used; the remote link position \(q_r(t)\) is intentionally omitted due to the difficulty of measuring link-side states in flexible-joint systems.

Let \(T_l(t)\) and \(T_r(t)\) denote the time-varying communication delays on the forward (local-to-remote) and backward (remote-to-local) channels, respectively, with known upper bounds \(T_l^\star\) and \(T_r^\star\). The delayed local link position received by the remote side is \(q_l(t - T_l(t))\), and the delayed remote joint position received by the local side is \(\theta_r(t - T_r(t))\).

The resulting \(P\!+\!d\) control laws for the local torque \(\tau_l\) and remote torque \(\tau_r\) are:
\begin{equation}
    \tau_{l} = K_{l} \bigl[\, q_{l}(t) - \theta_{r}(t - T_{r}(t)) \,\bigr] + B_{l} \dot{q}_{l}(t) 
    \label{eq:local_control}
\end{equation}
\begin{equation}
    \tau_{r} = -K_{r} \bigl[\, \theta_{r}(t) - q_{l}(t - T_{l}(t)) \,\bigr] - B_{r} \dot{\theta}_{r}(t) 
    \label{eq:remote_control}
\end{equation}
In these control laws, the proportional gains $K_{l}$ and $K_{r}$ correct position errors by applying force proportional to how far the system is from its target. The damping gains $B_{l}$ and $B_{r}$ add resistance to motion, smoothing movement and reducing oscillations. All gain terms are assumed to be positive constants. 

\subsection{DRL Agents for Adaptive Gain Tuning}
    Conventional controllers often depends on fixed gains from linearized models or manual adjustment, causing these methods to struggle with system nonlinearities, uncertain parameters, or dynamics that change over time. DRL directly addresses these limitations by adapting the controller gains online without any prior model. As a model-free approach, DRL eliminates the need for system identification, offline tuning, or expert knowledge. Fundamentally, DRL operates as a decision-making agent that learns an optimal policy(a mapping from states of the system to actions) by maximizing a cumulative reward signal through trial and error. It is naturally suited for continuous control tasks like gain tuning, as it can directly output real-valued control actions (the gains) and learn stable control policies purely through environmental interaction. By optimizing rewards based on closed-loop performance (e.g., minimizing tracking and damping error), the DRL agent continuously refines the control parameters in real time. This ensures robust control performance even when facing disturbances, setpoint changes, or shifts in operating conditions, making it particularly effective for complex or uncertain systems where traditional methods are inadequate\cite{b25}\cite{b35}.

\subsection{DRL Problem Formulation}

To implement a DRL agent for adaptive gain tuning, the control problem is formalized within the standard Markov Decision Process (MDP) framework. This involves defining the core components: the state representation, the action space, and the reward function.  
To simplify the training process, the DRL agent is implemented exclusively on the remote controller, while local control is omitted in this study and left for future work.

\begin{itemize}
    \item \textbf{State ($\mathbf{s}_t$)}: The state vector offers the agent a complete observation of the system’s dynamic condition at each time step $t$ . It is formulated according to the teleoperator dynamics as follows:
\begin{equation}
\label{eq:state_vector}
\begin{split}
    \mathbf{s}_t = [ & \theta_r(t), \dot{\theta}_r(t), q_l(t-T_{l}(t)), \\
                     & e_{\text{tracking}}(t), \int_{0}^{t} e_{\text{tracking}}(\tau)\, d\tau, \\
                     & e_{\text{damping}}(t), \int_{0}^{t} e_{\text{damping}}(\tau)\, d\tau ]^{T} ,
\end{split}
\end{equation}
where $\theta_r$ and $\dot{\theta}_r$ denote the remote joint (motor) position and velocity, respectively; $q_l(t - T_l(t))$ represents the delayed local link position;
$e_{\text{tracking}} = \theta_r - q_l(t - T_l(t))$ is the tracking error between the remote and delayed local positions; $\int e_{\text{tracking}}, d\tau$ is its integral;
$e_{\text{damping}} = \theta_r - q_l$ represents the flexibility (damping) error between the remote link position ($q_r$) and the remote joint (motor) position; and $\int e_{\text{damping}}, d\tau$ denotes its integral.

\item \textbf{Action ($\mathbf{a}_t$)}:  
The action $\mathbf{a}_t$ represents the output of the policy network, which determines the adaptive gain tuning for the remote controller. Specifically, the action defines the proportional gain ($K$) and damping injection gain vector ($B$) for each degree of freedom (DOF). It is expressed as
\begin{equation}
\mathbf{a}_t = [K_r^i, B_r^i], \quad i = \{1, 2, \ldots, n\},
\end{equation}
where $K_r^i$ and $B_r^i$ denote the proportional and damping gains of the $i_\text{th}$ joint of the remote manipulator, and $n$ is the total number of joints (motors).  
The gains of the local manipulator are kept constant and are not adaptively tuned.

\begin{figure*}[!t]
    \centering
    \begin{minipage}[t]{0.65\textwidth}
        \centering
        \includegraphics[width=0.99\textwidth]{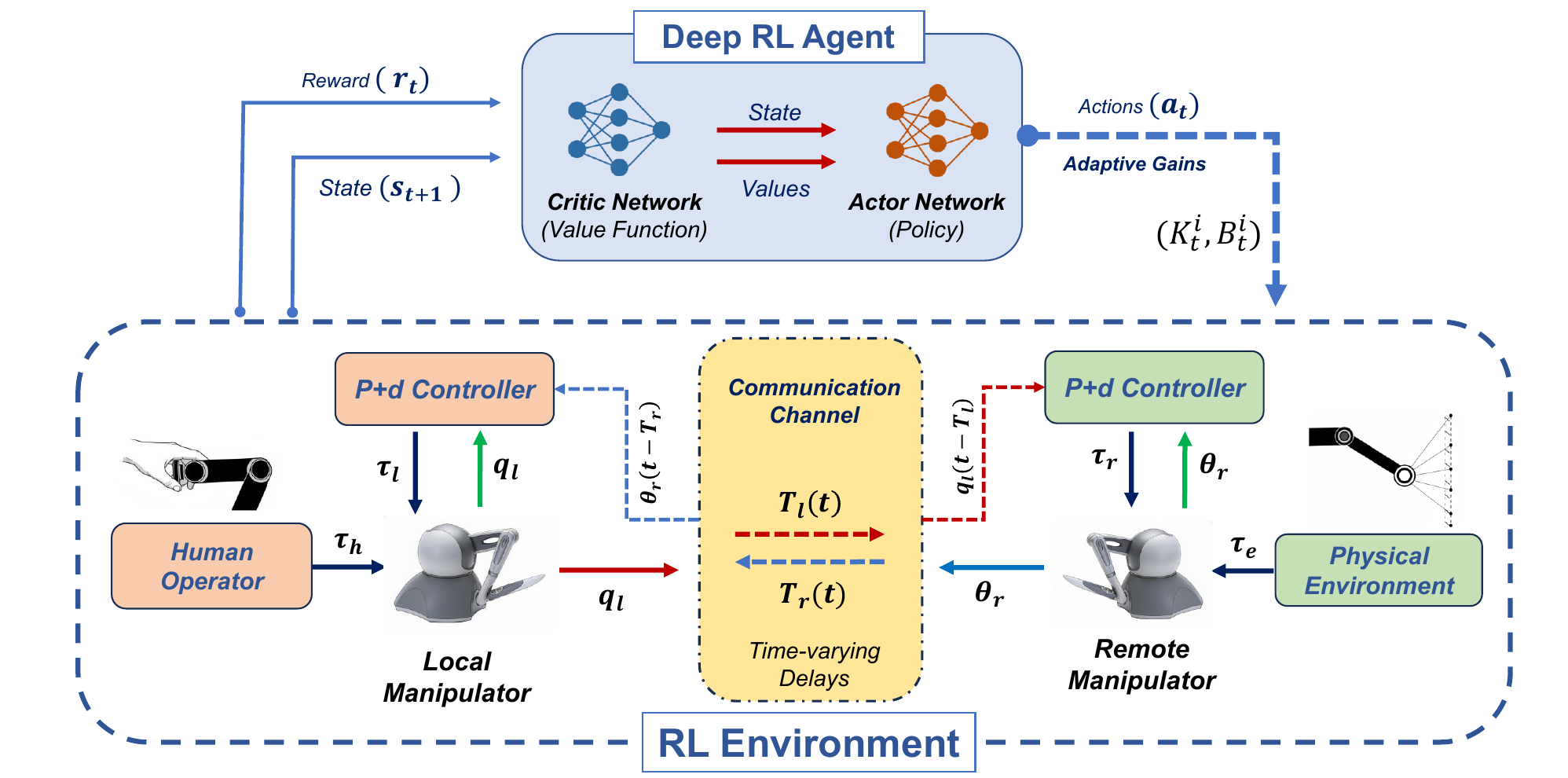}\\[2pt]
        \makebox[\textwidth]{\subcaption{(a)}}
        \label{fig:env}
    \end{minipage}
    \hfill
    \begin{minipage}[t]{0.34\textwidth}
        \centering
        \includegraphics[width=0.8\textwidth]{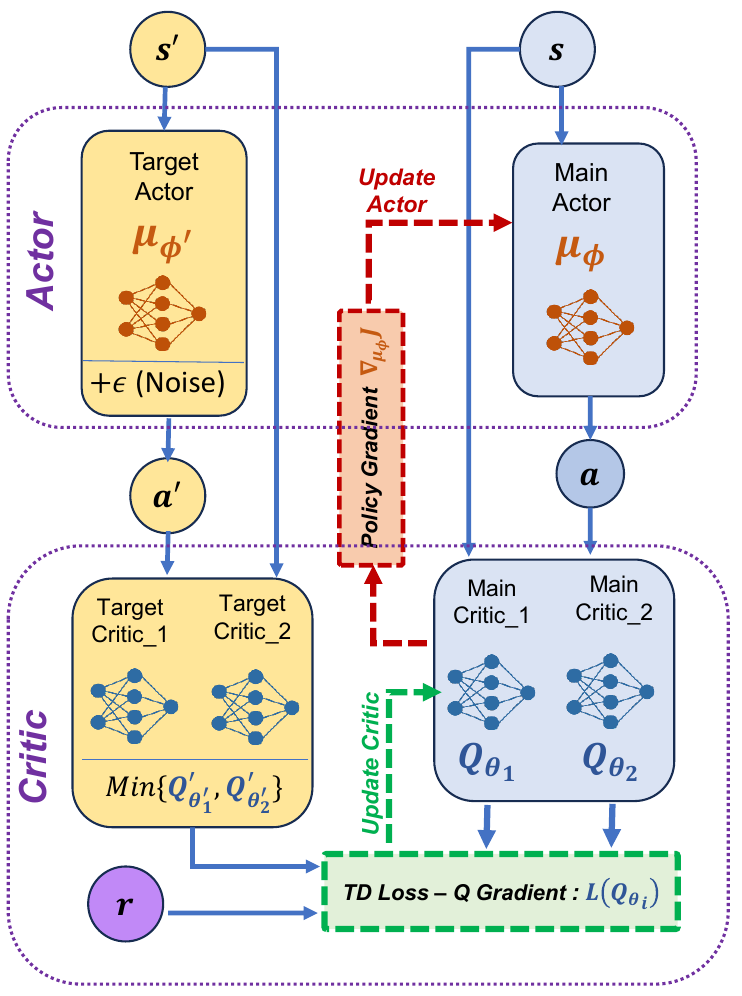}\\[2pt]
        \makebox[\textwidth]{\subcaption{(b)}}
        \label{fig:TD3}
    \end{minipage}

    \caption{Overview of the proposed control framework: (a) bilateral teleoperation system with joint flexibility and time-varying delays, and (b) TD3 network architecture used by the DRL agent for adaptive gain tuning.}
    \label{fig:combined}
\end{figure*}

\item \textbf{Reward ($r_t$)}:  
The reward function quantifies the control performance and guides the agent toward the desired behavior. The reward at time step $t$ is defined as
\begin{equation}
\label{eq:reward_function}
r_t = - \Big( 
\alpha \, \| e_\text{tracking} \|^2 
+ \beta \, \| e_\text{damping} \|^2 
+ \gamma \, \| \mathbf{a}_t \|^2 
\Big) ,
\end{equation}
where $\alpha$, $\beta$, and $\gamma$ are positive weighting coefficients that balance the contribution of each term.  
This formulation penalizes large tracking and damping errors through the first two terms, while $\| \mathbf{a}_t \|^2$ penalizes excessive changes in the controller gains to promote smooth and stable tuning.  
\end{itemize}

By interacting with the bilateral teleoperation environment as illustrated in Fig.~\ref{fig:combined}(a) , the DRL agent observes the system state $\mathbf{s}_t$, takes an action $\mathbf{a}_t = [K_r^i, B_r^i]$ to adaptively adjust the proportional and damping gains of the remote controller, and receives a reward $r_t$ based on the resulting tracking and damping performance.

\subsection{Twin Delayed Deep Deterministic Policy Gradient (TD3)}

The Twin Delayed Deep Deterministic Policy Gradient (TD3) \cite{b34} algorithm is a model-free and off-policy deep reinforcement learning method that extends the Deep Deterministic Policy Gradient (DDPG) framework for continuous control\cite{b33}. Its architecture comprises six neural networks: an online and target Actor, along with two distinct pairs of online and target Critics. TD3 improves upon DDPG by addressing its systematic issue of Q-value overestimation through three key mechanisms. First, it employs twin Critic networks, utilizing the minimum of their Q-value estimates to reduce overestimation bias. Second, it delays Actor updates, ensuring that policy improvements are guided by a more stable value function. Third, it applies target policy smoothing via additive, clipped Gaussian noise to target actions during Critic updates, which regularizes the policy and mitigates the exploitation of value estimation errors. The structure of the TD3 algorithm can be found in Fig. \ref{fig:combined}(b).
\section{Stability Analysis}

We consider a rigid local manipulator and a remote manipulator with joint flexibility (only on the remote side). 
Let the link and joint states be $q_l$, $q_r$, and $\theta_r$. 
The communication delays on the forward and backward channels are time-varying and bounded:
\begin{equation}
0 \le T_i(t) \le T_i^{\ast}, 
\qquad i \in \{l, r\}.
\label{eq:delay_bounds}
\end{equation}

Human and environment operators are modeled as passive mappings from velocity to torque:
\begin{equation}
\begin{aligned}
E_h &= - \int_0^t \dot{q}_l^{\top}(\sigma)\,\tau_h(\sigma)\,d\sigma + \kappa_l \ge 0, \\
E_e &= \phantom{-} \int_0^t \dot{q}_r^{\top}(\sigma)\,\tau_e(\sigma)\,d\sigma + \kappa_r \ge 0.
\end{aligned}
\label{eq:passivity}
\end{equation}

The delayed coordination errors (for the variable-delay case) are defined as
\begin{equation}
\begin{aligned}
e_l &= q_l - \theta_r\big(t - T_r(t)\big), \\
e_r &= \theta_r - q_l\big(t - T_l(t)\big).
\end{aligned}
\label{eq:errors}
\end{equation}

The local and remote torques follow a P+d structure with time-varying gains provided by the TD3 agent:
\begin{equation}
\begin{aligned}
\tau_l &= K_l(t)\,e_l + B_l(t)\,\dot{q}_l, \\
\tau_r &= -K_r(t)\,e_r - B_r(t)\,\dot{\theta}_r.
\end{aligned}
\label{eq:controller}
\end{equation}

The TD3 outputs are constrained to an admissible set $\mathcal{S}$ to preserve stability:
\begin{equation}
K_i^{\min} \le K_i(t) \le K_i^{\max}, 
\qquad 
B_i^{\min} \le B_i(t) \le B_i^{\max}.
\end{equation}

The delay-robustness inequality holds for all $t$:
\begin{equation}
4\,B_l(t)\,B_r(t) \ge \big(T_l^{\ast} + T_r^{\ast}\big)^2\,K_l(t)\,K_r(t) + \varepsilon,
  \varepsilon > 0,
\label{eq:robust_condition}
\end{equation}
where $\varepsilon$ provides a safety margin.

\subsection{Lyapunov--Krasovskii Functional}

We adopt the same Lyapunov--Krasovskii functional as in the paper by Nuno \cite{b23}, using constant positive weights independent of time-varying gains:
\begin{equation}
V = V_l
    + \eta\, V_r 
    + E_h + \eta E_e 
    + \frac{\eta K_l}{2}\, |q_l - \theta_r| ,
\label{eq:V}
\end{equation}

\begin{equation}
\begin{split}
V_r(q_r,\dot q_r,\theta_r,\dot\theta_r)
= \frac12\big( \dot q_r^{\top} M_r(q_r)\dot q_r
+ \dot\theta_r^{\top} J_r \dot\theta_r \big) \\
\quad + \frac12 (q_r-\theta_r)^{\top} S_r (q_r-\theta_r).
\end{split}
\end{equation}

\begin{equation}
V_l(q_l,\dot q_l)
= \frac{1}{2}\, \dot q_l^{\top} M_l(q_l)\, \dot q_l ,
\end{equation}

where $V_l$, $V_r$ represent the kinetic and potential energies of the local rigid and remote flexible subsystems, $\eta>0$ is a fixed scalar, and ${K}_l$ is a known lower bound. 

For variable delays, the following integral inequality, stated as Lemma 1 in \cite{b23}, is repeatedly used to prove stability:
\begin{equation}
-\!\!\int_0^t x^{\top}(\sigma)
    \!\!\int_{-T(\sigma)}^{0} 
    y(\sigma+\theta)\,d\theta\,d\sigma
\;\le\;
\frac{\alpha}{2}\|x\|_2^2 
+ \frac{(T^{\ast})^2}{2\alpha}\|y\|_2^2.
\label{eq:lemma}
\end{equation}

Differentiating $V$ along \eqref{eq:controller} and applying \eqref{eq:lemma} yields
\begin{equation}
\dot{V} \le 
-\lambda_l\,\|\dot{q}_l\|^2 
-\lambda_r\,\|\dot{\theta}_r\|^2,
\label{eq:Vdot}
\end{equation}
where $\lambda_l, \lambda_r > 0$ exist if and only if \eqref{eq:robust_condition} holds.

Remark: Since $V$ uses fixed positive weights, no $\dot{K}$ or $\dot{B}$ terms appear in $\dot{V}$. 
Hence, TD3 may vary the gains arbitrarily fast as long as they remain inside $\mathcal{S}$.

\section{Numerical Simulation \& Results}

The numerical simulations were conducted to evaluate the effectiveness of the proposed adaptive DRL control strategy under challenging teleoperation conditions involving joint flexibility, environmental uncertainties, and time-varying communication delays. All simulations were implemented in MATLAB/Simulink over a duration of $T_f = 40$~s with a sampling time of $T_s = 0.01$~s.

\subsection{Bilateral Teleoperation Setup}
The system comprises a rigid 2-DOF master (local) manipulator and a flexible-joint 2-DOF slave (remote) manipulator, with dynamics detailed in Section II. Joint flexibility on the remote side is modeled via link positions $q_r$ and motor positions $\theta_r$, coupled by a diagonal stiffness matrix $S_r = 100I$~N·m. Physical parameters include link lengths $l_1 = l_2 = 0.38$~m, link masses $m_{1l} = m_{1r} = 0.5$~kg and $m_{2l} = m_{2r} = 0.35$~kg, and remote motor inertia $J_r = 0.3I$~kg·m$^2$. The local controller uses fixed gains $K_l = 25$~N·m and $B_l = 10$~N·m·s.

\begin{figure}[htbp]
    \centering
    \includegraphics[width=0.95\linewidth]{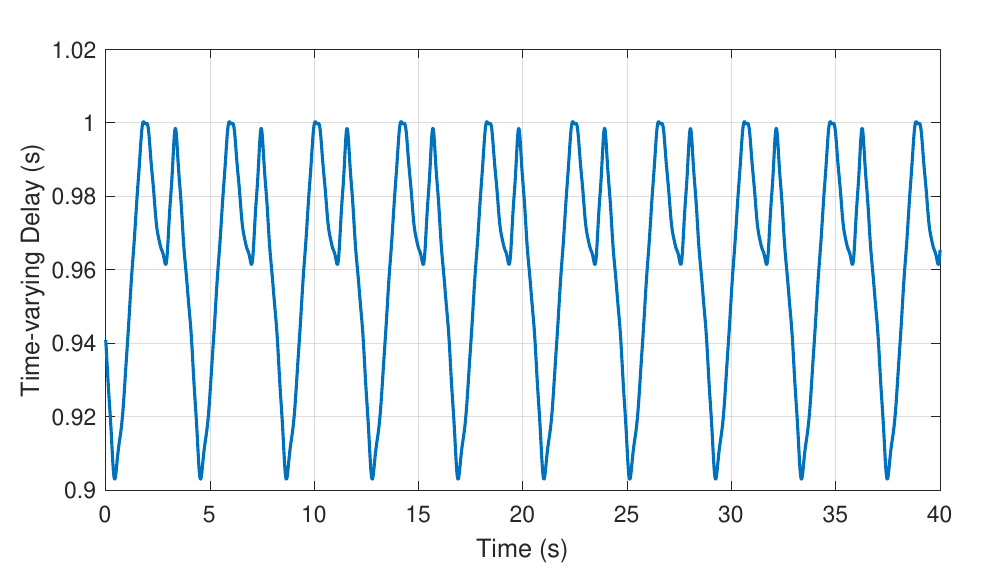}
    \caption{Communication channel delay profile.}
    \label{fig:delay}
\end{figure}

Time-varying communication delays $T_i(t)$ are bounded by $T_l^\star = T_r^\star = 1$~s (see Fig.~\ref{fig:delay}). The human operator is modeled as $\tau_h = K_s(q_h - q_l) - K_d \dot{q}_l$, with $K_s = 10$~N/m, $K_d = 2$~N·s/m, and $q_h$ a smooth sinusoidal reference. The environment torque is set to $\tau_e = 0$, representing free-space operation.

\subsection{DRL Training Details}
The TD3 agent was trained within a simulated bilateral teleoperation environment to adaptively tune the proportional and damping gains of the remote controller. The actor and critic networks were implemented as fully connected multilayer perceptrons with ReLU activations in the hidden layers. The actor’s final layer used a Softplus activation function to ensure all adaptive gains remained positive and smooth. Each critic network received both state and action inputs, which were processed through parallel feature-extraction layers before being merged to estimate the Q-value. Two critic networks were employed to reduce value overestimation and stabilize learning.

The main hyperparameters were set as follows: the discount factor $\gamma = 0.99$, minibatch size of $512$, and learning rates of $1\times10^{-4}$ for both the actor and critics, using the Adam optimizer with gradient clipping of~1. Gaussian exploration noise was applied with an initial standard deviation of~4, decaying exponentially to a minimum of~0.05 at a rate of $1\times10^{-7}$. The target networks were softly updated with coefficient $\tau = 0.005$. 

\begin{figure}[htbp]
    \centering
    \begin{minipage}[t]{0.48\columnwidth}
        \centering
        \includegraphics[width=\linewidth]{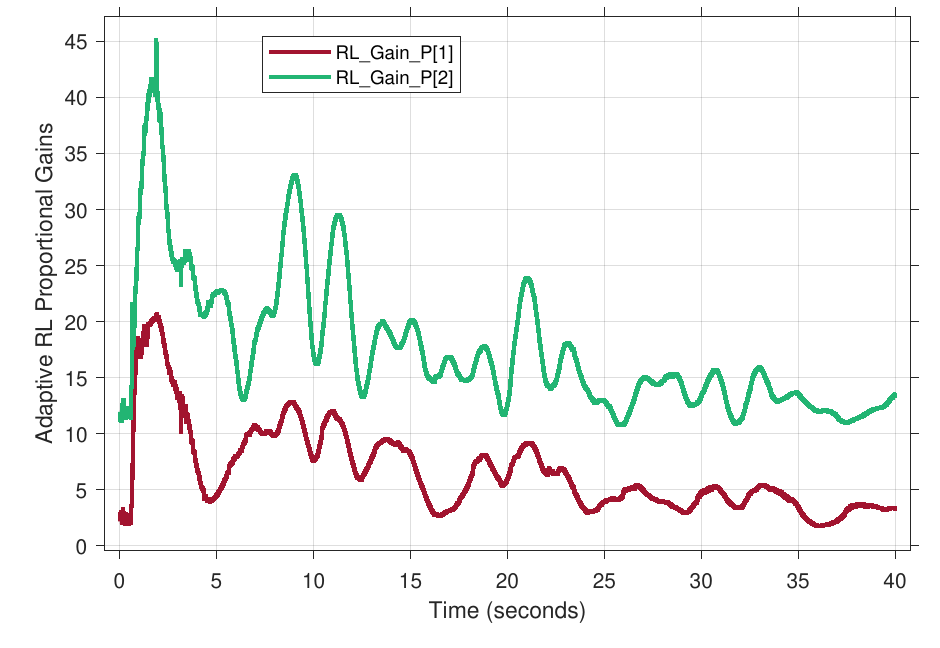}
        \label{fig:RL_P}
    \end{minipage}
    \hfill
    \begin{minipage}[t]{0.48\columnwidth}
        \centering
        \includegraphics[width=\linewidth]{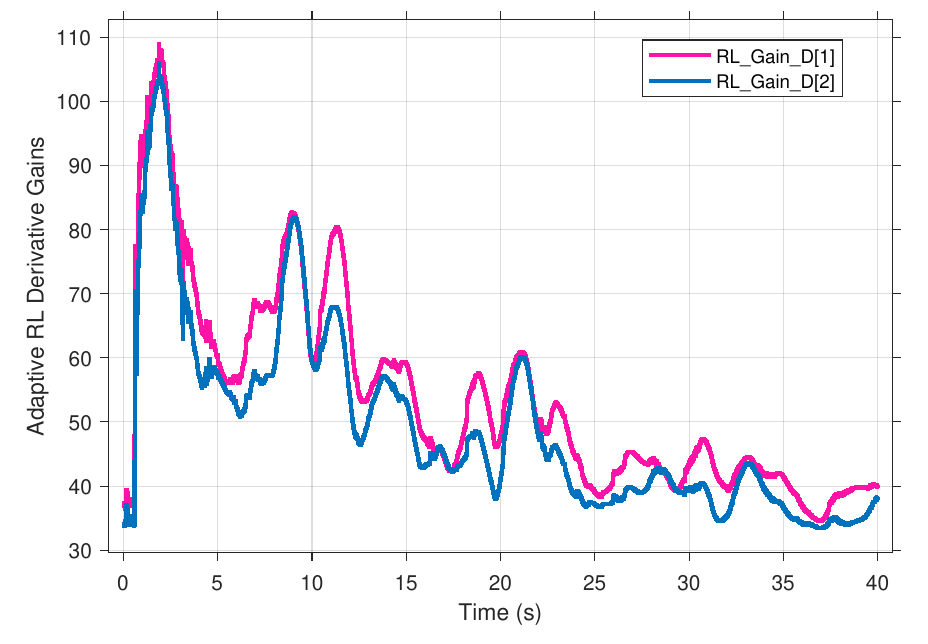}
        \label{fig:RL_D}
    \end{minipage}
    \caption{DRL-generated adaptive gains for $i=1,2$: (left) proportional gains $K_t^i$, and (right) derivative gains $B_t^i$.}
    \label{fig:RL_gains_combined}
\end{figure}

\subsection{Control Performance Evaluation}
The proposed DRL-based adaptive gain tuning controller demonstrates robust performance in stabilizing the flexible-joint teleoperation system while achieving high tracking accuracy, even in the presence of joint flexibility and time-varying communication delays.

\begin{figure}[htbp]
    \centering
    \includegraphics[width=0.95\linewidth]{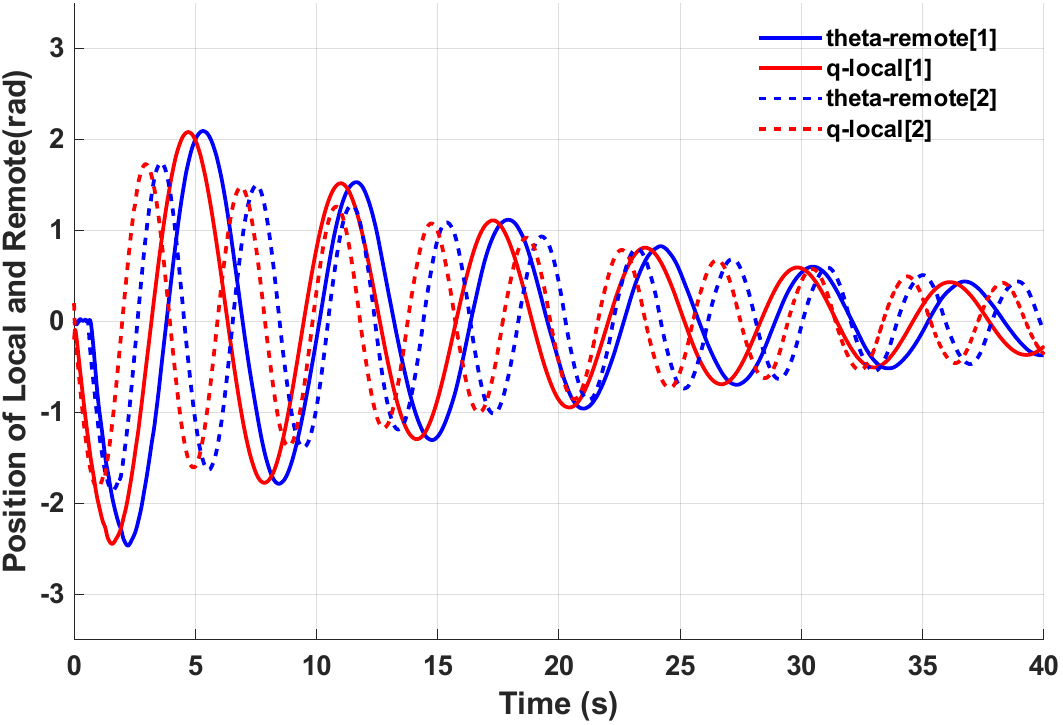}
    \caption{Time response of remote joint positions $\theta_r^i(t)$ and local link positions $q_l^i(t - T_l)$ for $i = 1, 2$.}
    \label{fig:positions}
\end{figure}

As shown in Fig.~\ref{fig:positions}, the time-domain response of the system reveals that the DRL controller effectively suppresses the oscillatory behavior inherent in flexible-joint systems with delays. Both the remote joint positions ($\theta_r$) and the delayed local link positions ($q_l$) rapidly converge to a synchronized steady state, indicating strong stabilization capabilities.

\begin{figure}[htbp]
    \centering
    \includegraphics[width=0.95\columnwidth]{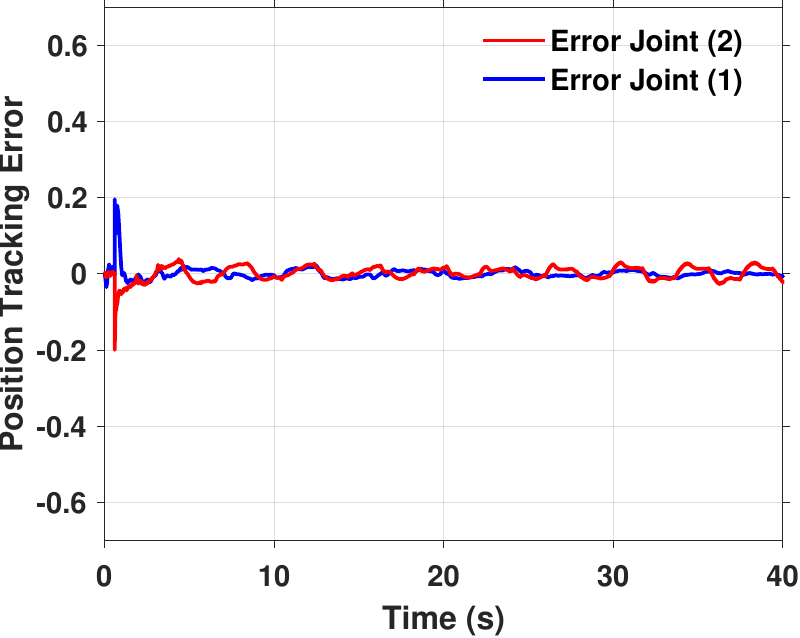}
    \caption{Tracking error between the delayed local link position $q_l(t - T_{l})$ and the remote joint angle $\theta_r(t)$.}
    \label{fig:error_tracking}
\end{figure}

Fig.~\ref{fig:error_tracking} presents the position tracking error, defined as $\theta_r - q_l(t - T_l)$, where $T_l$ denotes the forward communication delay. For both joints, the tracking error converges swiftly to a neighborhood of zero, underscoring the controller’s ability to maintain tight coordination between the master and slave sides despite time-varying delays. The negligible steady-state error further attests to the efficacy of the DRL framework in compensating for such uncertainties.

\begin{figure}[htbp]
    \centering
    \includegraphics[width=0.95\columnwidth]{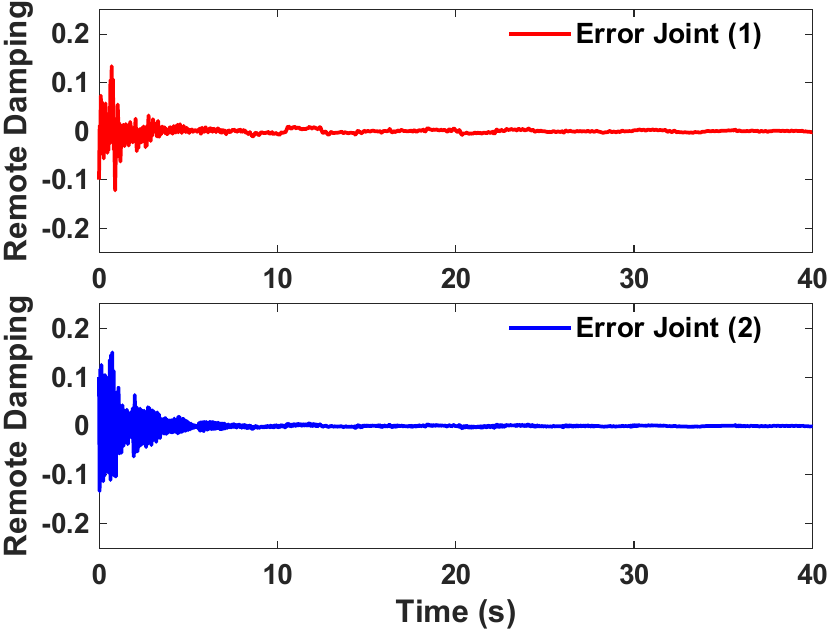}
    \caption{Remote link-joint damping error ($\theta_r - q_r$)}
    \label{fig:error_damping}
\end{figure}

Moreover, Fig.~\ref{fig:error_damping} illustrates the remote link--joint damping error ($\theta_r - q_r$), which reflects the level of vibration induced by joint flexibility. The rapid attenuation of this error for both joints confirms that the adaptive gain tuning mechanism successfully emulates virtual damping, thereby mitigating the destabilizing influence of joint elasticity.

In summary, the experimental results validate that the proposed DRL-based adaptive controller achieves fast stabilization, accurate position tracking, and effective vibration suppression in a flexible-joint teleoperation system, successfully addressing the dual challenges posed by mechanical compliance and time-varying network delays.


\section{Conclusion and Future Work}
This work presented a hybrid control framework that combines a stable P+d controller with a TD3-based DRL agent to address joint flexibility and time-varying delays in bilateral teleoperation. The DRL agent adaptively tunes the remote-side gains in real time, while Lyapunov–Krasovskii analysis guarantees stability under bounded delays. Simulations on a 2-DOF flexible manipulator with delays up to \SI{1}{s} confirmed that the approach achieves fast stabilization, accurate position tracking, and effective vibration suppression, successfully handling the dual challenges of joint flexibility and uncertain network delays.
Future work will extend adaptive gain tuning to the master side, incorporate impedance adaptation for interactions with stiff environments, explore a multi-agent framework \cite{sheykh} to coordinate independent reinforcement learning on both master and slave sides, validate the approach on a physical teleoperation platform under real-world conditions, and integrate sim-to-real techniques \cite{icrom_landing}.


\begin{thebibliography}{00}

\bibitem{b1} Y. Yang, Y. Yang, X. Li, and C. Hua, “Adaptive synchronization control of multimanipulator teleoperation system under constrained discrete-time network communication,” \textit{Int. J. Robust Nonlinear Control}, vol. 33, no. 3, pp. 1807--1820, Feb. 2023.

\bibitem{b2} M. Babaiasl, S. N. Goldar, M. H. Barhaghtalab, and V. Meigoli, “Sliding mode control of an exoskeleton robot for use in upper-limb rehabilitation,” in \textit{Proc. 3rd RSI Int. Conf. Robot. Mechatronics (ICROM)}, Oct. 2015, pp. 694--701.

\bibitem{b3} A. Rashvand, M. J. Ahmadi, M. Motaharifar, M. Tavakoli, and H. D. Taghirad, “Adaptive robust impedance control of haptic systems for skill transfer,” in \textit{Proc. 9th RSI Int. Conf. Robot. Mechatronics (ICRoM)}, Nov. 2021, pp. 586--591.

\bibitem{b4} H. J. Shin, H. K. Yoo, J. H. Lee, S. R. Lee, K. Jeong, and H. S. Moon, “Robotic single-port surgery using the da Vinci SP\textsuperscript{\tiny\textregistered} surgical system for benign gynecologic disease: A preliminary report,” \textit{Taiwan. J. Obstet. Gynecol.}, vol. 59, no. 2, pp. 243--247, Mar. 2020.

\bibitem{b5} F. Gaboardi, G. Pini, N. Suardi, F. Montorsi, G. Passaretti, and S. Smelzo, “Robotic laparoendoscopic single-site radical prostatectomy (R-LESS-RP) with da Vinci Single-Site\textsuperscript{\tiny\textregistered} platform: Concept and evolution of the technique following an IDEAL phase 1,” \textit{J. Robot. Surg.}, vol. 13, no. 2, pp. 215--226, Apr. 2019.

\bibitem{b6} P. F. Hokayem and M. W. Spong, “Bilateral teleoperation: An historical survey,” \textit{Automatica}, vol. 42, no. 12, pp. 2035--2057, Dec. 2006.

\bibitem{b7} R. J. Anderson and M. W. Spong, “Bilateral control of teleoperators with time delay,” in \textit{Proc. IEEE Int. Conf. Syst., Man, Cybern.}, Aug. 1988, vol. 1, pp. 131--138.

\bibitem{b8} N. Chopra, M. W. Spong, R. Ortega, and N. E. Barabanov, “On tracking performance in bilateral teleoperation,” \textit{IEEE Trans. Robot.}, vol. 22, no. 4, pp. 861--866, Aug. 2006.

\bibitem{b9} N. Chopra, M. W. Spong, and R. Lozano, “Synchronization of bilateral teleoperators with time delay,” \textit{Automatica}, vol. 44, no. 8, pp. 2142--2148, Aug. 2008.

\bibitem{b10} E. Nuño, R. Ortega, and L. Basañez, “An adaptive controller for nonlinear teleoperators,” \textit{Automatica}, vol. 46, no. 1, pp. 155--159, Jan. 2010.

\bibitem{b32} S. Ganjefar, H. Momeni, and F. Janabi-Sharifi, “Teleoperation systems design using augmented wave-variables and Smith predictor method for reducing time-delay effects,” in \textit{Proc. IEEE Int. Symp. Intell. Control}, 2002, pp. 333--338.

\bibitem{b11} D. B. Venkateswaran and Z. Qu, “A passivity-shortage based control design for teleoperation with time-varying delays,” \textit{IEEE Robot. Autom. Lett.}, vol. 5, no. 3, pp. 4070--4077, Apr. 2020.

\bibitem{b12} S. Lu, Y. Ban, X. Zhang, B. Yang, S. Liu, L. Yin, and W. Zheng, “Adaptive control of time delay teleoperation system with uncertain dynamics,” \textit{Front. Neurorobot.}, vol. 16, p. 928863, Jul. 2022.

\bibitem{b13} J. Wang, J. Tian, X. Zhang, B. Yang, S. Liu, L. Yin, and W. Zheng, “Control of time delay force feedback teleoperation system with finite time convergence,” \textit{Front. Neurorobot.}, vol. 16, p. 877069, May 2022.

\bibitem{b14} B. Yang, C. Liu, L. Zhang, L. Teng, J. Tian, S. Xu, and W. Zheng, “Novel design of three-channel bilateral teleoperation with communication delay using wave variable compensators,” \textit{Electronics}, vol. 14, no. 13, p. 2595, Jun. 2025.

\bibitem{b15} C. Uyulan, “Robust passivity-based nonlinear controller design for bilateral teleoperation system under variable time delay and variable load disturbance,” \textit{Nonlinear Eng.}, vol. 13, no. 1, p. 20220358, Jan. 2024.

\bibitem{b16} Y. H. Chang, C. Y. Yang, and H. W. Lin, “Robust adaptive-sliding-mode control for teleoperation systems with time-varying delays and uncertainties,” \textit{Robotics}, vol. 13, no. 6, p. 89, Jun. 2024.

\bibitem{b17} K. Chen and H. Zhang, “Design of synchronization tracking adaptive control for bilateral teleoperation system with time-varying delays,” \textit{Sensors}, vol. 22, no. 20, p. 7798, Oct. 2022.

\bibitem{b18} H. Chen and Z. Liu, “Time-delay prediction-based Smith predictive control for space teleoperation,” \textit{J. Guid. Control Dyn.}, vol. 44, no. 4, pp. 872--879, Apr. 2021.

\bibitem{b19} M. Ebrahimian, M. Pourmokhtari, M. Ghiyasi, B. Yazdankhoo, and B. Beigzadeh, “Online bilateral predictive control for time-delayed teleoperation of snake-like robots,” \textit{J. Intell. Robot. Syst.}, vol. 110, no. 2, p. 80, May 2024.

\bibitem{b20} P. N. Dao and H. A. Duc, “Nonlinear RISE based integral reinforcement learning algorithms for perturbed bilateral teleoperators with variable time delay,” \textit{Neurocomputing}, vol. 605, p. 128355, Nov. 2024.

\bibitem{b21} K. Patiño, E. Slawiñski, M. Moran-Armenta, V. Mut, F. G. Rossomando, and J. Moreno-Valenzuela, “Neural networks in the delayed teleoperation of a skid-steering robot,” \textit{Mathematics}, vol. 13, no. 13, p. 2071, Jun. 2025.

\bibitem{b22} B. Xia, X. Tian, B. Yuan, C. Yang, Z. Li, B. Liang, and X. Wang, “Agent-based space teleoperation: Mitigating time delays with deep reinforcement learning,” \textit{IEEE Trans. Syst., Man, Cybern.: Syst.}, Jul. 2025.

\bibitem{b31} A. Khanzadeh, S. Ganjefar, and M. Ghaemifar, “An adaptive controller for cooperative teleoperation system with time-varying delay and formation,” in \textit{Proc. 11th Int. Conf. Robot. Mechatronics (ICRoM)}, Dec. 2023, pp. 48--53.

\bibitem{b23} E. Nuño, I. Sarras, L. Basañez, and M. Kinnaert, “Control of teleoperators with joint flexibility, uncertain parameters and time-delays,” \textit{Robot. Auton. Syst.}, vol. 62, no. 12, pp. 1691--1701, Dec. 2014.

\bibitem{b24} J. Xu, H. Li, and Q. Zhang, “Adaptive control for circulating cooling water system using deep reinforcement learning,” \textit{PLoS One}, vol. 19, no. 7, p. e0307767, Jul. 2024.

\bibitem{b25} F. Yin, X. Yuan, Z. Ma, and X. Xu, “Vector control of PMSM using TD3 reinforcement learning algorithm,” \textit{Algorithms}, vol. 16, no. 9, p. 404, Aug. 2023.

\bibitem{b26} A. H. Barjini, S. A. Kolagar, S. Yaqubi, and J. Mattila, “Deep reinforcement learning-based motion planning and PDE control for flexible manipulators,” \textit{IEEE Robot. Autom. Lett.}, Jul. 2025.

\bibitem{b27} P. Tomei, “A simple PD controller for robots with elastic joints,” \textit{IEEE Trans. Autom. Control}, vol. 36, no. 10, pp. 1208--1213, Oct. 2002.

\bibitem{b28} R. Kelly and V. Santibáñez, “Global regulation of elastic joint robots based on energy shaping,” \textit{IEEE Trans. Autom. Control}, vol. 43, no. 10, pp. 1451--1456, Oct. 2002.

\bibitem{b29} E. Nuño, L. Basañez, R. Ortega, and M. W. Spong, “Position tracking for nonlinear teleoperators with variable time delay,” \textit{Int. J. Robot. Res.}, vol. 28, no. 7, pp. 895--910, Jul. 2009.

\bibitem{b35} X. Yuan, Y. Wang, R. Zhang, Q. Gao, Z. Zhou, R. Zhou, and F. Yin, “Reinforcement learning control of hydraulic servo system based on TD3 algorithm,” \textit{Machines}, vol. 10, no. 12, p. 1244, Dec. 2022.

\bibitem{b34} S. Fujimoto, H. van Hoof, and D. Meger, “Addressing function approximation error in actor-critic methods,” in \textit{Proc. Int. Conf. Mach. Learn. (ICML)}, Jul. 2018, pp. 1587--1596.

\bibitem{b33} T. P. Lillicrap, J. J. Hunt, A. Pritzel, N. Heess, T. Erez, Y. Tassa, D. Silver, and D. Wierstra, “Continuous control with deep reinforcement learning,” \textit{arXiv:1509.02971}, Sep. 2015.

\bibitem{sheykh} M. Sheykh, H. A. Talebi, and I. Sharifi, “A centralized adaptive PID control of telerehabilitation systems using multi-agent systems theory,” in \textit{Proc. 2024 32nd Int. Conf. Electr. Eng. (ICEE)}, 2024, pp. 1--7.

\bibitem{icrom_landing} K. Abbasloo, A. Attarzadeh, M. P. Sangdeh, H. R. Ahouei, and S. Shamaghdari, “Sim-to-real deep Q-learning with human-aided demonstrations for vision-based precision landing of multirotors,” in \textit{Proc. 2024 12th RSI Int. Conf. Robot. Mechatronics (ICRoM)}, 2024, pp. 131--136.

\bibitem{b30} M. Farahmandrad, S. Ganjefar, H. A. Talebi, and M. Bayati, “A novel cooperative teleoperation framework for nonlinear time-delayed single-master/multi-slave system,” \textit{Robotica}, vol. 38, no. 3, pp. 475--492, 2020.

\end{thebibliography}
\end{document}